\documentclass[11pt,a4paper,twoside]{article}
\usepackage{epsfig}
\usepackage{baltlat5}
\usepackage{wrapfig}
\usepackage[authoryear]{natbib}
\begin{document}
\ \
\vspace{-0.5mm}

\setcounter{page}{1}
\vspace{-2mm}

\titleb{SENENMUT: AN ANCIENT EGYPTIAN ASTRONOMER}

\begin{authorl}
\authorb{Bojan Novakovi{\' c}}{}
\end{authorl}

\begin{addressl}
\addressb{}{Astronomical Observatory, Volgina 7, 11160 Belgrade 74, Serbia}
\end{addressl}

\begin{summary}
The celestial phenomenon have always been a source of wonder and
interest to people, even as long ago as the ancient Egyptians.
While the ancient Egyptians did not know all the things about
astronomy that we do now, they had a good understanding of the
some celestial phenomenon. The achievements in astronomy of
ancient Egyptians are relatively well known, but we know very
little about the people who made these achievements. The goal of
this paper is to bring some light on the life of Senenmut, the
chief architect and astronomer during the reign of Queen
Hatshepsut.
\end{summary}




\sectionb{1}{INTRODUCTION}

As early as several thousand years ago people were interested in
astronomy and they have had some knowledge about celestial
phenomenon. This kind of interest existed in almost all ancient
civilization, although it raised from different purposes and
motivations. The ancient Egyptians were interested in astronomy,
mainly for practical and religious purposes.

This paper is devoted to achievements of ancient Egyptians
astronomy (section 2.) and to the life of an ancient Egyptians
astronomer Senenmut (section 3.).

\vskip1mm

\sectionb{2}{ASTRONOMY IN ANCIENT EGYPT}

Astronomy was very important to the ancient Egyptians and played a
different part for that people than it does in many cultures.
Although their achievements were far less advanced than those of
some other ancient civilization, many of them are very important
and deserve our attention. Here, we will mention some of them.

The invention of the 365-days calendar, based on astronomical
observation. The development of this type of calendar probably
took place at least as far back as 2,000 B.C., but the first
calendar developed in Egypt was lunar calendar and it was
developed about 3000 B.C - Mankind's first measurement of time.
The beginning of the Egyptian year was declared when there was a
flood, as they noticed that the flood begins with the star Sirius,
the brightest star in the sky. This incident represented the
beginning of the agricultural year in Egypt. The year had 365 days
divided into 12 months and each month had 30 days. They made the
remaining five days feast days, called the Epagomenal Days, or the
days upon the year, and added them at the end of the year. Months
of the year were divided into three seasons, namely: the flood
season, the planting season, and the harvest season. The year, the
season, the month and the day in which the king assumed power was
usually recorded by the Egyptians in their documents.

The development of instruments of quantitative astronomical
measurement. These included the sundial, water clocks, and the
merkhet. The ancient Egyptians used instruments or indicators for
observing the circumpolar stars. They would then draw a
north-south axis line on the ground marking its direction, which
was required for the proper orientation of important building
projects. One of the instruments used was called "Merkhet,"
(similar to an astrolabe), which could mean "indicator." It
consisted of a horizontal, narrow wooden bar with a hole near one
end, through which the astronomer would look to fix the position
of the star. The other instrument, called the "bay en imy unut,"
or palm rib, had a V-shaped slot cut in the wider end through
which the priest in charge of the hours looked to fix the star.

Telling time by the stars. Astronomy in ancient Egypt was the best
way to tell the time during the night. They recognized a number of
constellations and other groups of stars. These groups of stars,
called decans, were used for telling time at night. Each group of
stars rose forty minutes later each night. Observing the position
of a group of stars in relation to the day of the year would tell
a person what time it was. Theoretically, there were 18 decans,
however, due to dusk and twilight only twelve were taken into
account when reckoning time at night. Since winter is longer than
summer the first and last decans were assigned longer hours.
Tables to help make these computations have been found on the
inside of coffin lids. The columns in the tables cover a year at
ten day intervals. The decans are placed in the order in which
they arise and in the next column, the second decan becomes the
first and so on.

The achievements in astronomy of ancient Egyptians also include:

\begin{itemize}
    \item {Knowledge of stellar constellations - at least 43 constellations
were familiar to the Egyptians in the 13th century B.C.}
    \item {Knowledge of planetary astronomy - five planets were known to the
Egyptians; the retrograde motion of Mars was known; the revolution
of Mercury and Venus around the Sun was known.}
    \item {Astronomy was also used in positioning the pyramids. They are
aligned very accurately, the eastern and western sides run almost
due north and the southern and northern sides run almost due
west.}
\end{itemize}

\sectionb{3}{SENENMUT}

The earliest known star maps in Egypt are found as a main part of
a decor in a tomb (TT 353) at Thebes on the West bank of the Nile
(e.g. Leser 2006). The tomb was build during of the Egyptian 18th
dynasty, and it belonged to Queen Hatshepsut's vizer and calendar
registrar Senenmut (also known as Senmut or Senemut).

But, who really was Senenmut? Senenmut was of low birth, born to
literate provincial class parents, Ramose and Hatnofer. Despite
his non-royal origin Senenmut was given more prestigious titles
and became high steward of the king. There is no doubt that much
more is known about Senenmut than any other non-royal Egyptian.

The list of Senenmut's titles are very long, but the first of all
he was an architect, government official and tutor of Neferure -
Queen Hatshepsut's daughter. Senenmut originally entered the royal
court during the reign of Tuthmosis II, under Hatshepsut he would
eventually hold over 80 titles \citep{Dor88} during his period as
an official and administrator working in the royal court. Other
dimensions of his career are suggested by the presence of an
astronomical ceiling in his tomb at Deir el Bahari and about 150
ostraca in his tomb at Qurna, including several drawings, as well
as lists, calculations, various reports and literary works. No
doubt the workmen were instructed to decorate his tomb with items
of interest in the life of Senenmut.

The social classification of the family has also been a central
point of the discussion. Probably at that time about 5\% of the
population was able of reading and writing. Therefore, Tyldesley
(1996) placed the family in the "upper" social class, which
mastered these stages of civilization, because in her opinion
Senenmut would not have been able to start successfully into his
career without these abilities. In this connection is also
unclear, how or where Senenmut has started his career. Able to
read and write he could have started his career as a low civil
servant. However, it is also possible that he had started with a
military career and then changed into the administration. As far
as we know it was quite usual that retiring officers were awarded
with an administrative position. The destroyed inscriptions in his
monument, TT71, which contain text fragments possibly give some
information about the beginning of his career.

Beside the offices mentioned above, which he surely executed, he
also got numerous "courtly titles" - like the one called "Only
friend of the Pharao". These titles most likely testify the
extraordinary confidence of Hatshepsut.

Concerning the end of Senenmut there are more speculations than
facts. At least until regnal year 16 of Hatshepsut/Thutmosis III.
he held his offices. Apparently thereafter, his tracks are lost.
His unfinished monument, TT353, was closed, some his figures
therein, and also in TT71, were destroyed. There is no information
that he had been buried in one of his tombs.

The astronomical ceiling in Senenmut's tomb (TT 353) is divided
into two sections representing the northern and the southern
skies. The southern (Figure 1.) is decorated with a list of
decanal stars, as well as constellations of the southern sky
belonging to it like Orion and Canis Major. Furthermore, the
planets Jupiter, Saturn, Mercury and Venus are shown and
associated deities who are travelling in small boats over the sky.
Thus, the southern ceiling marks the hours of the night. The
northern shows constellations of the northern sky with the large
bear (Ursa major) in the center. The other constellations could
not be identified. On the right and left of it there are 8 or 4
circles shown and below them several deities each carrying a sun
disk towards the center of the picture. The inscriptions
associated with the circles mark the original monthly celebrations
in the lunar calendar, whereas the deities mark the original days
of the lunar month \citep{Mey82}.

The map on the southern panel proves to reflect a specific
conjunction of planets around the longitude of Sirius. The four
planets Jupiter, Saturn, Mercury and Venus are relatively easily
recognizable. The planet Mars is not included in the actual
grouping and at first sight seems to be missing in the map.
However, Mars is also pictured in the Senenmut map, but it is
represented by an empty boat in the west. This seems to refer to
the fact that Mars was retrograde so that in this backward
movement (well known phenomenon to the Egyptians) the Mars
position was perhaps not consider to be "concrete". Using these
facts, Egyptologist were able to date that this particular
configuration of planets occurred in the sky in 1534 BC
\citep{vanS00}.

Modern chronologists tend to agree that Hatshepsut reigned as
pharaoh from 1479 to 1458 BC, but there is no definitive proof of
the beginning date. Some other sources proposed that Hatshepsut
could have assumed power as early as 1512 BC. Consequently, it is
not clear whether or not the celestial phenomenon, mentioned
above, was happened within the lifetime of Senenmut.

\begin{figure}
\includegraphics[height=7.4cm]{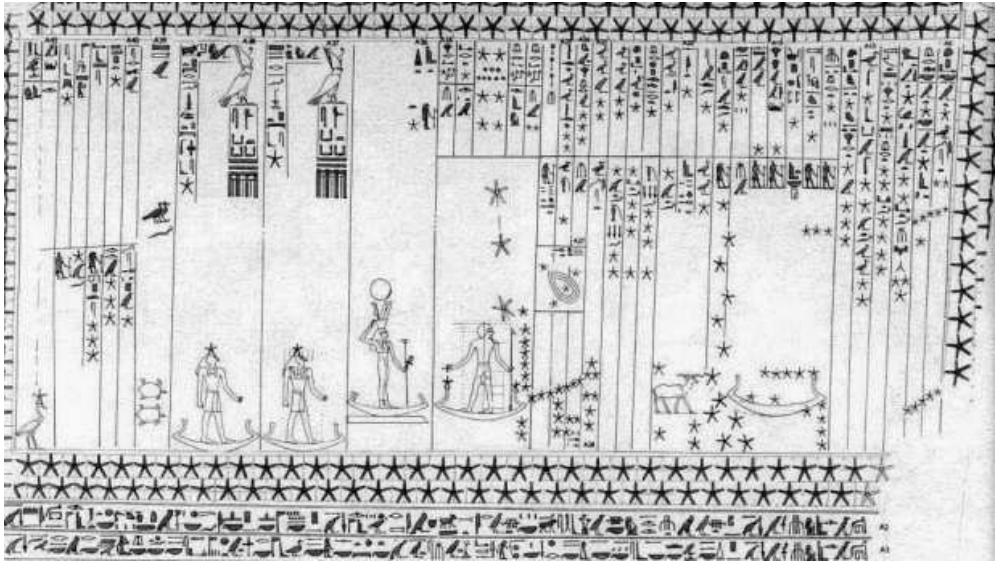}
\caption{The southern part of the astronomical ceiling in
Senenmut's tomb (TT 353)}
\end{figure}

\vskip 3mm

\sectionb{4}{CONCLUSIONS}

A short review of the achievements in astronomy of ancient
Egyptians, presented here, indicates that Egyptian astronomy
deserves more attention. Probably, there are a lot of things
waiting to be discovered about their astronomy.

The available evidence about the life of Senenmut suggests that he
was an astronomer. Although, it may conflict with some other
results (e.g. Shaw 2003), the obvious probability exists that when
the rare conjunction occurred in 1534 BC it was within the
lifetime of Senenmut. However in order to answer this question a
further investigation is necessary.





\end{document}